\documentclass[12pt]{article}  
\usepackage[dvips]{graphicx}
\usepackage{amsmath,amssymb,amsbsy}
\usepackage{amsfonts}
\usepackage{epsfig}
\newcommand{\beq}[1]{ \begin{equation}\label{#1}}
\newcommand{\eeq}{\end{equation}}
\newcommand{\ba}{\begin{array}}

\newcommand{\beqa}{\begin{eqnarray}}
\newcommand{\eeqa}{\end{eqnarray}}
\newcommand{\bear}{\begin{array}{c}}
\newcommand{\bearr}{\begin{array}{cc}}
\newcommand{\ear}{\end{array}}

\newcommand{\mbf}{\mathbf}

\newcommand{\R}{I\kern-.3em{R}}

\begin{document}\thispagestyle{empty}
\null
~~~~~~~~~~~~~~~~~~~~~~~~~~~~~~~~~~~~~~~~~~~~~~~~~~~~~~CERN-PH-TH-2014-077
\begin{center}
\vskip 1.cm
{\Large\bf{Do we understand near-forward elastic scattering\\ 

up to TeV energies? \footnote{Contribution to the special issue of the International Journal of Modern Physics A on "Elastic and diffractive scattering" coordinated by Christophe Royon}
}}
\vskip 1.cm
{\bf Claude Bourrely}
\vskip 0.3cm 
 Aix-Marseille Universit\'e,
D\'epartement de Physique,\\ Facult\'e des Sciences, site de Luminy,
13288 Marseille, Cedex 09, France\\
\vskip 0.5cm
{\bf Jacques Soffer}
\vskip 0.3cm
        Department of Physics, Temple University, Philadelphia, PA 19122-6082, USA\\
\vskip 0.5cm
{\bf Tai Tsun WU}
\vskip 0.3cm
Gordon McKay Laboratory
Harvard University Cambridge, MA 02138, USA\\ and\\
Theoretical Physics Division, CERN, 1211 Geneva 23, Switzerland\\
\vskip 2.5cm

\begin{center}
{\bf Abstract}
\end{center}
\end{center}

\vskip 0.5cm

In 1970, on purely theoretical grounds, all total hadronic total cross
sections were predicted to increase without limit for higher and higher
energies.  This was contrary to the conventional belief at that time.  In
1978, an accurate phenomenological model was formulated for the case of
proton-proton and antiproton-proton interactions.  The parameters for this
model were slightly improved in 1984 using the additional available
experimental data.  Since then, for thirty years these parameters have not
changed.  This development, including especially the difficult task of
formulating this phenomenological model and the comparison of the
predictions of this model with later experimental results, is summarized.

\vskip 0.5cm

\vskip 0.5cm
\clearpage
\newpage
\section{Introduction}
\label{intro}
In the 1958 International Conference on High Energy Physics,
Oppenheimer \cite{OPP} gave the concluding remarks.  He said, among other
points:

"There are areas where we know very little -- extremely high energy\\
\vskip-0.5cm
collisions, for instance -- where little can be done by anyone."

Physics is basically an experimental science.  At that time over half a
century ago when Oppenheimer gave his speech, there were no high-energy
accelerator where the collisions can be called even remotely "extremely
high energy", in the sense that the incident particles are extremely
relativistic in the center-of-mass system.

This situation began to change about a decade later: the 200-GeV proton
accelerator was to be built at National Accelerator Laboratory, now called
the Fermilab, and the Intersecting Storage Ring (ISR) was to be built at
CERN.  At the laboratory energy of 200 GeV, the center-of-mass energy of
each proton is about ten times the proton mass (times the velocity of
light squared), which is extremely relativistic. The proton energy of ISR
is even higher.  That these accelerators would be soon available gave
motivation, indeed urgency, to study theoretically the Oppenheimer
challenge of understanding extremely high energy collisions.

A main role of theoretical physics is to give quantitative predictions
that can be either confirmed or refuted by later experiments.  In this way
our understanding of physics can be deepened.

What quantity should be studied first at such extremely high energies?  At
any energy, the overall property of a collision process is provided by the
total cross section, which is the sum of the integrated cross sections of
all possible scattering and production processes.  In turn, the cross
sections of these various processes are necessarily affected by the
behavior of the total cross section.

Motivated by the Oppenheimer challenge and the accelerators to be
available soon, we embarked on the task of studying theoretically the
proton-proton total cross section at the energies of these accelerators
and beyond.

In the nineteen sixties, there were two distinct schools of thought on the
high-energy scattering of strongly interacting particles: the droplet
model of Yang and collaborators \cite{ttwu3} and the Regge pole model \cite{regge}.  The
droplet model has numerous successes for many processes at high energies,
and is based on the following observations from experimental data: the
elastic differential cross sections appear to approach limiting values as
the incident energy $E \to \infty$, above about 300 MeV excitation energy, the
nucleon has many excited states and the large-angle proton-proton elastic
cross section drops spectacularly with energy.  In contrast, the Regge
pole model deals mostly with scattering processes where some quantum
number is exchanged.  The intuition obtained from these two models guided
our thinking on total cross sections at high energies.

Shortly after Professor Yang developed the droplet model for hadron-hadron
collisions, the question was asked: What are the experimental evidences
for or against the existence of limiting values for total cross section?
By studying the experimental data on the ratio

$$       \rho = \frac{\mbox{Real part of pp} \rightarrow  \mbox{pp in the forward direction}}
              {\mbox{Imaginary part of pp} \rightarrow   \mbox{pp in the forward direction}}$$
in the case of proton-proton elastic scattering and using dispersion
relation, there was a weak and preliminary indication that there may not be
such a limiting value in this case.  This weak indication actually played
an important role in our decision in the nineteen sixties to study the
total cross section at high energies.  This episode has been described in \cite{ttwu4} and we will come back to it in Section 3. In the above expression, the amplitude for $pp \to pp$ does not include the contribution from the Coulomb interaction, (See Eq. (\ref{eq:nc}) in Section 3).

For the purpose of this study, we began by writing down a list of the
basic features of the interaction between elementary particles, features
that are valid not only at high energies but at all energies:\\
- three spatial dimensions and one time dimension\\
- relativistic kinematics\\
- unitarity and\\
- particle production.\\
Even though each of these four features may be considered to be "trivial",
it is nevertheless not easy to have a model with all these four features.
The simplest way to have all features is to have a relativistic quantum
field theory.

As emphasized by Yang and Mills \cite{YM1,YM2}, "... the usual principle of
invariance under isotopic spin rotation is not consistent with the concept
of localized fields."  This inconsistency is avoided in theories with a
gauge invariance of the second kind.  For this reason, the decision was
made to study the high-energy behavior of a relativistic quantum gauge
field theory in 3+1 dimensions.  At that time in the nineteen sixties,
there was no choice because the only such theory that was understood was
the Abelian case.  It has turned out that this was most
fortunate: even now nearly half a century later, the high-energy behavior
of non-Abelian Yang-Mills theory remains unknown, even for the simplest
case with the gauge group {\it SU(2)}.

     The theoretical investigation of the high-energy behavior, with fixed
transverse momentum, for this Abelian relativistic quantum gauge
field theory began in the late nineteen sixties.  The main
theoretical result, obtained in 1970, is that the total cross section
increases without limit for higher and higher energies; this result is
then interpreted to imply that the total cross sections of
strongly interaction particles all have to increase this way \cite{ttwu1}.  This is
in agreement with the preliminary indication from the
experimental measurements on the ratio $\rho$ mentioned above.
     This theoretical development, together with some of the early
phenomenology has been describe in detail \cite{ttwu2}.  This book \cite{ttwu2} is
already over quarter of a century old; reference \cite{ttwu9} gives a later summary
of the situation.

\section{Basic Idea of Phenomenology}
\label{sec2}
It is an interesting and important theoretical statement that hadronic
total cross sections all increase without bound for higher and higher
energies.  Nevertheless, such a statement by itself is not of great help
for us to gain a better understanding of nature.

It is the foremost purpose of theoretical physics to make quantitative
predictions.  The most important predictions are those that open up new
directions of investigation and are verified accurately by subsequent
experiments; they allow us to have a novel way of understanding nature.
The next type of important predictions are those that open up new
directions but are refuted by subsequent experiments; they tell us that
the novel view needs suitable modification \footnote{To the best of the knowledge of the authors, this wisdom is due to Niels Bohr}.

In the present context of scattering processes at extremely high energies,
the step after the asymptotic calculation of the perturbation series for
the Abelian gauge theory is therefore to develop a quantitatively accurate
phenomenology.  What does this mean?  Since the internal structure of a
proton is exceedingly complicated and not understood quantitatively from
first principles, this phenomenology must satisfy the following two
conditions:

On the one hand, the proton-proton and proton-antiproton total cross
sections must increase without bound at very high energies, and

On the other hand, this phenomenology must give a reasonable description
of these two cross sections at the relatively high-energy region of the
existing experimental data.

{\it The development of such a phenomenology requires a great deal of physical
intuition and is perhaps the most difficult step for the present approach}
to scattering at extremely high energies, an attempt to response to the
Oppenheimer challenge.  More precisely, the difficulty was as follows.
While the field-theoretic calculation \cite{ttwu2,ttwu1} leading to the prediction of
increasing total cross section played a central role, it was physical
intuition that determined which field theoretic results should be
incorporated into the phenomenology and which ones must be rejected.

Here are the two most important and difficult decisions for the development of
our phenomenology, the first one making use of one of the results from
field theory, while the second one purposely contradicting another result
from field theory.

(I) A choice must be made of the variable for describing the increasing
total cross sections.  For this purpose, the Regge language is most
helpful.  It was known that the leading Regge singularity should be above
1, but the question is: what is the nature of this singularity?  The
possible choices are:\\
- moving Regge pole, as in the usual Regge theory\\
- fixed Regge pole\\
- moving Regge cut or\\
- fixed Regge cut.

Our initial inclination was to follow the prevailing thinking in the
nineteen sixties and to use a moving Regge pole.  However, our intuitive
feeling was that the leading singularity being a Regge pole is intimately
related to the nature of the underlying quantum field theory.  Regge poles
are obtained from the summation of the ladder diagrams in $\phi^3$ theory,
which is a super renormaliable field theory.  In general, this property of
being super renormalizable is connected with the ladder-like diagrams
being described by a Fredholm integral equation, and this is the origin of
the moving Regge pole.  In contrast, quantum gauge theories in four
dimensions are never super renormalizable, and the explicit calculation
for the Abelian case gives a fixed Regge cut.  This was our reason in 1978
to use a fixed Regge cut above 1 \cite{BSW79}.  The other two cases of a fixed
Regge pole and a moving Regge cut were considered too unnatural to be used
for our phenomenology.

(II) We also need to make a decision how to describe the absorption in the
scattering of strongly interacting particles.  The specific issue is: when
the impact distance between the two incident particles is small, what is
the amount of absorption that should be incorporated into the
phenomenology?

In the field-theoretic calculation for the quantum Abelian gauge field
theory, this absorption is only partial even at extremely high energies.
Since this lack of complete absorption has a rather complicated origin, a
major issue was whether this property was likely to hold for the
interactions between strongly interacting particles.  This was an
agonizing choice, and we finally decided to go against this field-theoretic
result and take the absorption in this limit to be complete.  The basic
reason for this choice was that it seemed to be against physical intuition
that, in the language of Yang and collaborators, some of the "stuff" is
absorbed while others not \cite{ttwu3}.

It has been gratifying that our phenomenology has worked out well after
these two decisions based on our physical intuition.  While we have to
make a number of other choices to complete our phenomenology, the above
two are the most difficult and far-reaching ones.

To describe the experimental data taken at the relatively low energies available
to experiments forty years ago, a new model was proposed in 1978 \cite{BSW79}, including Regge backgrounds. Besides those pertaining to the Regge terms, there is a total of six
parameters for $pp$ and $\bar p p$ elastic scattering.  From the overall fit to
the existing data in 1978, the values of these six parameters \cite{BSW79} are
given in the left column of Table 1.

     Six years later, when there were significantly more experimental data
at high energies, the overall fit was repeated \cite{BSW84}.  The revised
values of these six parameters are given in the right column of Table
1. It should be emphasized that, in these six years from 1978 to
1984, the expressions used to describe the model is not altered at
all; these formulas are given explicitly later in this Section 2.
     It is interesting to compare the two columns of values in Table 1:
the six values have not changed much due to the additional
information.  This implies that this new model, sometimes referred to
as the BSW Model, is quite robust.  There has been no further change
of these parameter values in the thirty years since.

Both for the energies of the present-day colliders and for the purpose of studying
the asymptotic behavior of the model at high energies, all the Regge backgrounds
can be neglected. The BSW model is given by the
following matrix element for elastic scattering
\begin{equation}
  \label{eq:one}
  \mathcal{M}(s,\mbf{\Delta}) = \frac{is}{2\pi}\int d\mbf{x}_\perp
  e^{-i\mbf{\Delta}\cdot \mbf{x}_\perp}D(s,\mbf{x}_\perp)~,
\end{equation}
where $s$ is the square of the center-of-mass energy, $\mbf{\Delta}$ is the
momentum transfer, $\mbf{x}_\perp$ is the impact parameter and all
spin variables have been omitted. For this model we use for the opacity
\begin{equation}
D(s,\mbf{x}_\perp) = 1-e^{-\Omega(s,\mbf{x}^{2}_\perp)}~,
  \label{eq:opac}
\end{equation}
with
\begin{equation}
  \label{eq:omDef}
  \Omega(s,\mbf{x}^{2}_\perp)=\mathcal{S}(s)F(x^{2}_\perp)~,
\end{equation}
where $x_\perp \equiv \mbf{x}_\perp|$~.
The function $\mathcal{S}(s)$ is given by the complex symmetric expression, obtained from the high energy behavior
of quantum field theory \cite{ttwu2,ttwu1}
\begin{equation}
  \label{eq:Sdef}
  \mathcal{S}(s)= \frac{s^c}{(\ln s)^{c'}} + \frac{u^c}{(\ln u)^{c'}}~,
\end{equation}
with $s$ and $u$ in units of $\mbox{GeV}^2$, where $u$ is the
third Mandelstam variable \cite{SM}. In this Eq. (\ref{eq:Sdef}), $c$ and $c'$ are two 
dimensionless constants given in Table 1. That they
are constants implies that the Pomeron is a fixed Regge cut as discussed above. For the asymptotic behavior at high energy and modest momentum
transfers, we have to a good approximation
\begin{equation}
  \label{eq:uAp}
  \ln u = \ln s -i\pi~,
\end{equation}
so that
\begin{equation}
  \label{eq:S2}
    \mathcal{S}(s)= \frac{s^c}{(\ln s)^{c'}} + 
\frac{s^ce^{-i\pi c}}{(\ln s-i\pi)^{c'}}~.
\end{equation}
Because $F$ depends on $\mbf{x}_\perp$ only through $x^{2}_\perp$, the Fourier transform in Eq. (\ref{eq:one}) simplifies to
\begin{equation}
  \label{eq:M2nd}
  \mathcal{M}(s,\Delta) = is\int_0^\infty dx_\perp\,x_\perp\,J_0(x_\perp\Delta)
\left[1-e^{-\mathcal{S}(s)F(x^{2}_\perp)}\right]~,
\end{equation}
where $\Delta \equiv |\mbf{\Delta}|$.
The function $F(x_\perp)$ is taken to be related to the electromagnetic form factor
$G(t)$ of the proton, where $t= -\mbf{\Delta}^2$ is the Mandelstam variable for
the square of the momentum transfer.  Specifically, $F(x^{2}_\perp)$ is defined as in
\cite{BSW79} via its Fourier transform $\tilde{F}(t)$ by
\begin{equation}
  \label{eq:FtildDef}
  \tilde{F}(t) = f[G(t)]^2\frac{a^2+t}{a^2-t}~,
\end{equation}
with 
\begin{equation}
  \label{eq:Gdef}
  G(t)=\frac{1}{(1- t/m_1^2)(1- t /m_2^2)}~.
\end{equation}
The remaining four parameters of the model, $f$, $a$, $m_1$ $\mbox{and}$ $m_2$, are given in Table 1.\\
In the next section we will recall some of the early successes of our approach at the CERN $\bar {p} p$ collider
and at the FNAL Tevatron and the next section will be devoted to a preliminary discussion of the sitution at the Large Hadron Collider.  

We define the ratio of the real to imaginary parts of the forward amplitude, mentioned earlier in the introduction, 
$\rho(s) = \frac{\mbox{Re}~\mathcal{M}(s, t=0)}{\mbox{Im}~\mathcal{M}(s, t=0)}$,
 the total cross section 
$\sigma_{tot} (s) = (4\pi/s)\mbox{Im}~\mathcal{M}(s, t=0)$, 
the differential cross section
${d\sigma (s,t)/dt} = \frac{\pi}{s^2}|\mathcal{M}(s,t)|^2$, 
and the integrated elastic cross section 
$\sigma_{el} (s) = \int dt \frac{d\sigma (s,t)} { dt}$. One important feature of the BSW model is, as a consequence of Eq. (\ref{eq:S2}), the fact that the phase of the amplitude is built in. Therefore real and imaginary parts of the amplitude cannot be chosen independently and we will show now how to test them.

\begin{table}[htb]
\begin{center}
\begin{tabular}{|c|c|c|}
\hline
Year & 1979 & 1984\\
\hline\raisebox{0pt}[12pt][6pt]{$c$}&0.151    & 0.167     \\[4pt]
\hline\raisebox{0pt}[12pt][6pt]{$c'$}&0.756    & 0.748     \\[4pt]
\hline\raisebox{0pt}[12pt][6pt]{$m_1$}&0.619    & 0.586     \\[4pt]
\hline\raisebox{0pt}[12pt][6pt]{$m_2$}&1.587    & 1.704     \\[4pt]
\hline\raisebox{0pt}[12pt][6pt]{$f$}  &8.125    & 7.115     \\[4pt]
\hline\raisebox{0pt}[12pt][6pt]{a}    &2.257    & 1.953     \\[4pt]
\hline
\end{tabular}
\caption {Pomeron fitted parameters for $pp (\bar pp)$. Comparison of the 
1979 and 1984 solutions..}
\label{tab:table1}
\end{center}
\end{table}

\section{Early successes}
\begin{figure}[h]
\vspace*{-3.5ex}
\begin{center}
\includegraphics[width=9.0cm]{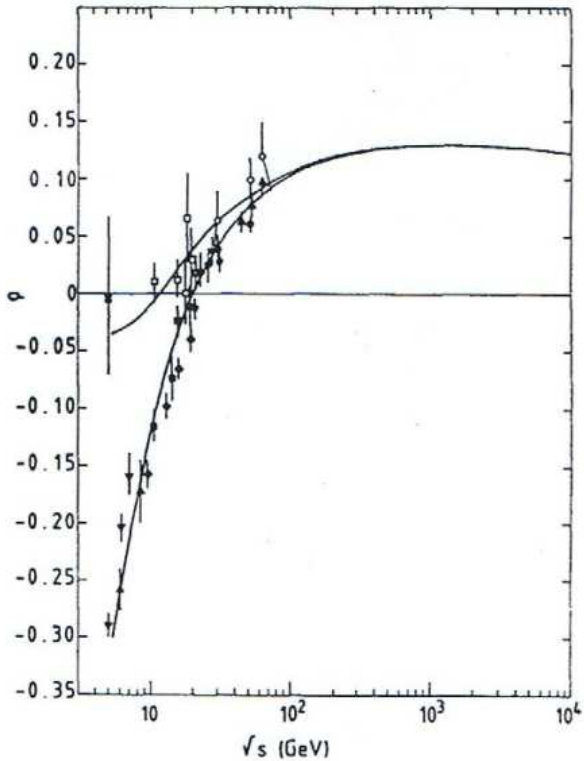}
\caption[*]{\baselineskip 1pt
 The ratio $\rho(s) = \frac{\mbox{Re}~\mathcal{M}(s, t=0)}{\mbox{Im}~\mathcal{M}(s, t=0)}$ for $pp$ (black points) and $\bar p p$ (open points)
elastic scattering and the BSW predictions up to high energies (Taken from Ref. \cite{BSW84}).}
\label{fig:rho}
\end{center}
\end{figure}

Before showing some early successes of our phenomenology we would like to
come back to an important observation which gave the first hint against 
the possibility that the total cross section would remain constant at very high energies.
This is indeed related to the behavior of $\rho(s)$ which is shown in Fig. \ref{fig:rho}. The forward scattering amplitude is expected to satisfy a dispersion relation. This means that the real part of this amplitude can be written as an integral over the imaginary part, which is essentially the total cross section. If the $pp$ total cross section does approach a finite limit, then this ratio $\rho(s)$ must approach zero at high energies. In the mid-sixties the experimentally measured values of $\rho(s)$ in the low energy region, below $\sqrt{s} \sim$ 10GeV, were negative and increasing toward zero. However, the rate of increase seemed to have a tendency to overshoot to become positive and this was the first indication for a possible increasing total cross section at high energies. Indeed positive $\rho(s)$ values were later obtained as displayed in Fig. 1, which also shows the BSW results and predictions up to high energies.
\vskip 1.cm
Let us now turn to some successful BSW predictions for near-forward $\bar p p$ elastic scattering at the FNAL and CERN colliders energies. For this kinematic region near the forward direction, one must consider the contribution of the Coulomb amplitude $a^C$, so Eq. (\ref{eq:one}) is replaced by
\begin{equation}
  \label{eq:nc}
\mathcal{M}(s,\mbf{\Delta}) \pm a_{\pm}^{C}(-\Delta^2),
\end{equation}
the upper sign is for $\bar p p$ while the lower one is for $pp$ scattering and
\begin{equation}
  \label{eq:ac}
	a_{\pm}^{C}(t) = 2\alpha s\frac{G^{2}(t)}{|t|}\mbox{exp}[\mp i\alpha\phi(t)],
\end{equation}	
where $\alpha$ is the fine structure constant, $\phi(t)$ is the West-Yennie phase \cite{wy} given by $\phi(t) = \mbox{ln}(t_0/|t|) - \gamma$, with
$t_0 = 0.08 \mbox{GeV}^2$ and $\gamma \sim 0.577$ is the Euler constant.	\\
At the FNAL-Tevatron, the E710 experiment running at $\sqrt{s}$=1.8TeV, has obtained $\sigma_{tot}=72.8 \pm 3.1$ mb and $\sigma_{el}/\sigma_{tot}= 0.23 \pm 0.012$ \cite{amos}, whereas the BSW predictions are 74.8 mb and 0.230 respectively. They were also able to extract the following $\rho$ value, $\rho = 0.140 \pm 0.069$ \cite{amos2}. This important measurement is in agreement with the BSW prediction, but has unfortunately little significance because of its large error. These data are reported in Fig. 2 together with the results of the CDF experiment at two different Tevatron energies $\sqrt{s}$=1.8TeV and $\sqrt{s}$=546GeV \cite{cdf}
and the results of UA(4) at the CERN $\bar p p$ collider at $\sqrt{s}$=541GeV \cite{augier2}. At $\sqrt{s}$=1.8TeV CDF found $\sigma_{tot}=80.03 \pm 2.24$ mb and $\sigma_{el}/\sigma_{tot}= 0.246 \pm 0.004$, at variance with the E710 results.
\begin{figure}[ht]
\vspace*{-3.5ex}
\begin{center}
\includegraphics[width=10.0cm]{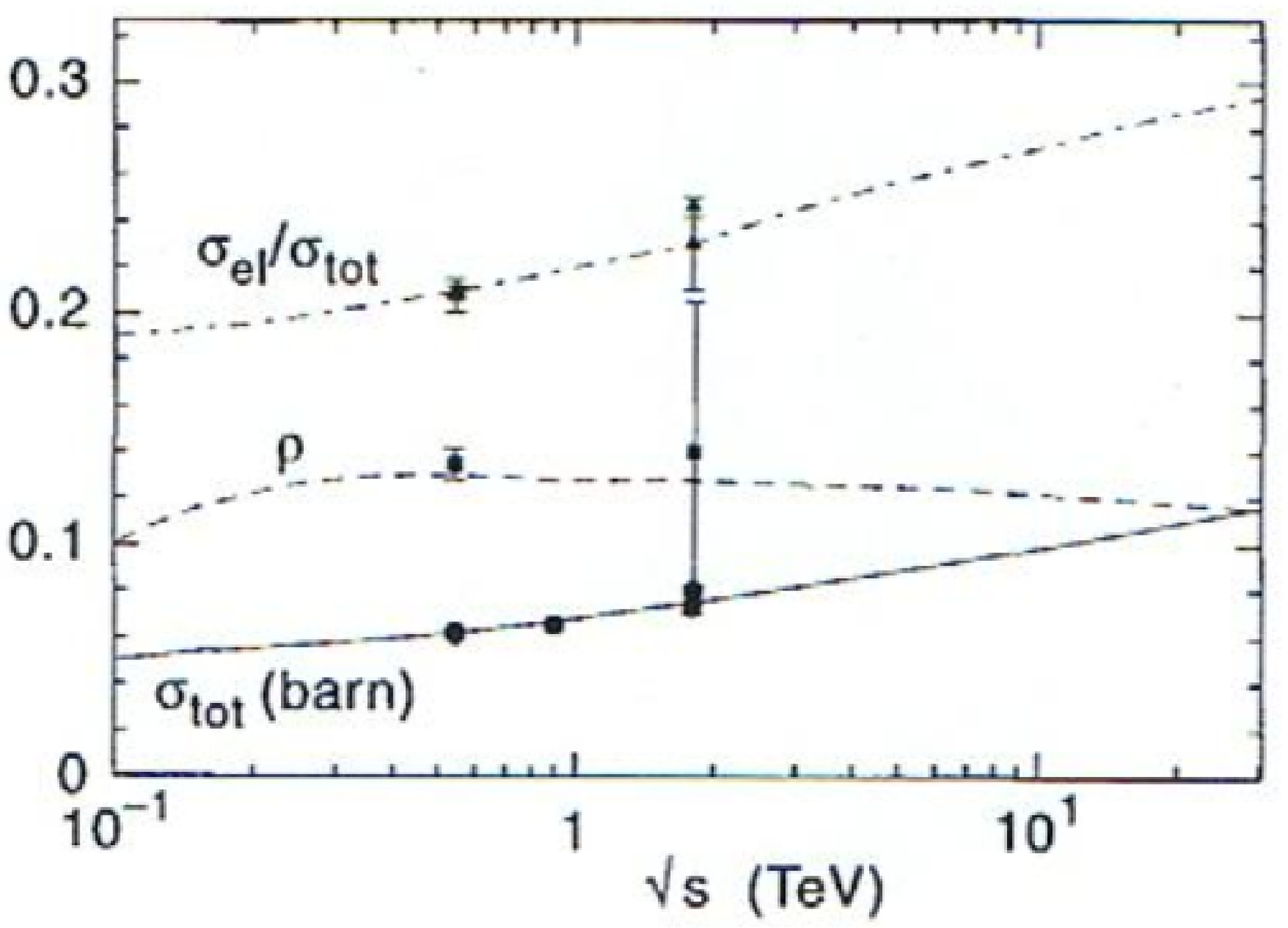}
\caption[*]{\baselineskip 1pt
Pre LHC era, data on the total cross section $\sigma_{tot}$ in barns, the ratio $\sigma_{el}/\sigma_{tot}$ of the integrated elastic cross section to the total cross section and the ratio $\rho(s)$, as function of $\sqrt{s}$, together with the BSW predictions from Ref. \cite{BSW84}.}
\label{fig:tot}
\includegraphics[width=10.0cm]{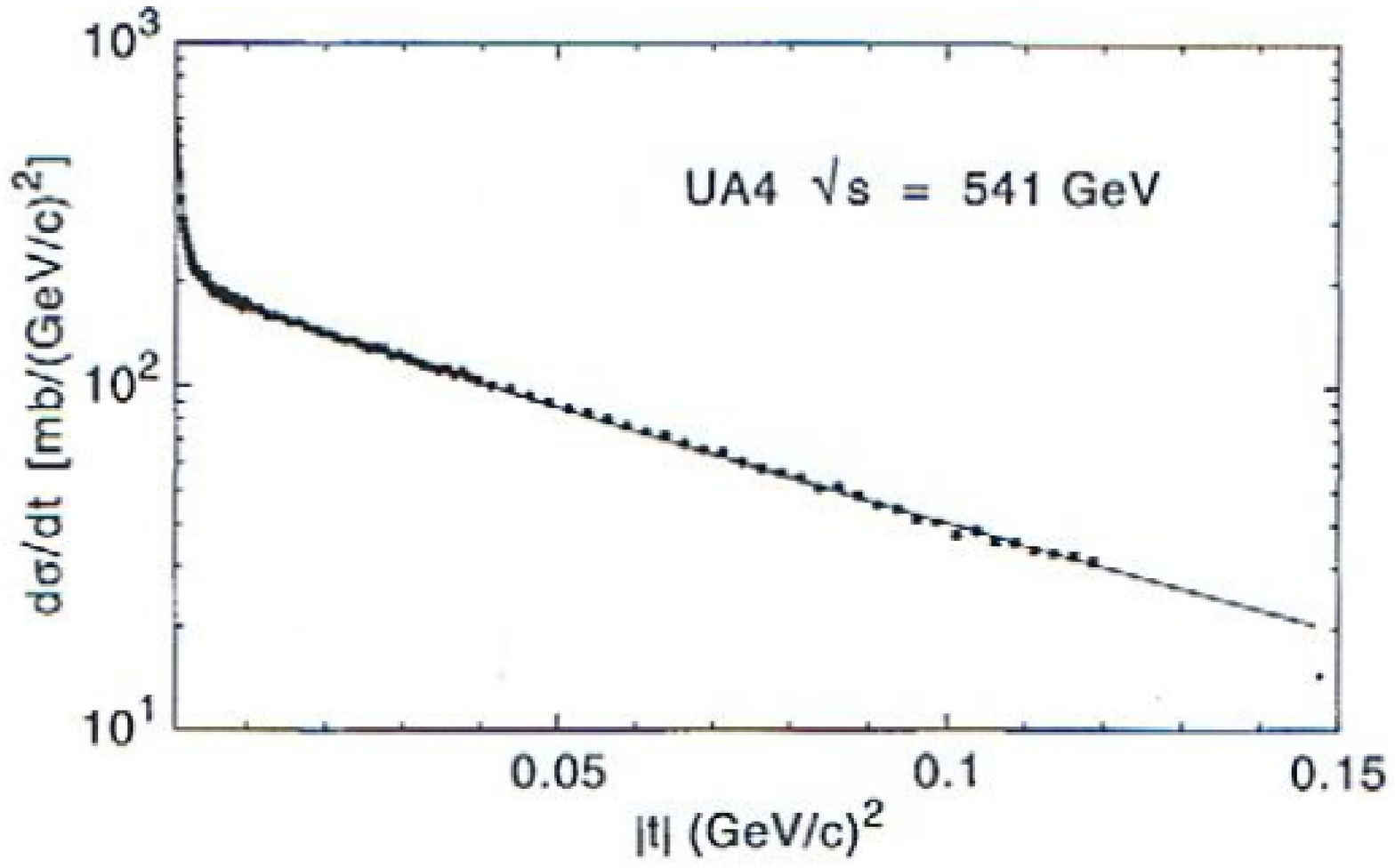}
\caption[*]{\baselineskip 1pt
 $d\sigma/dt$ for near-forward $\bar p p$ elastic scattering at $\sqrt{s}$= 541GeV data from Ref. \cite{augier}, the curve is the BSW prediction \cite{bsw87} (Taken from Ref. \cite{bsw93}).}
\label{fig:dsigua4}
\end{center}
\end{figure}
However at $\sqrt{s}$=546GeV, the CDF results $\sigma_{tot}=61.26 \pm 0.93$ mb and $\sigma_{el}/\sigma_{tot}=0.210 \pm 0.002$ agree well with those of UA(4), $\sigma_{tot}=63.0 \pm 2.1$ mb and $\sigma_{el}/\sigma_{tot}=0.208 \pm 0.007$. The UA(4) experiment has obtained a very precise value for
the parameter $\rho$, $\rho=0.135 \pm 0.015$, from the measurement of $d\sigma/dt$ in the Coulomb-nuclear interference region \cite{augier}, as shown
in Fig. \ref{fig:dsigua4}. One notices the rapid rise of the cross section in the very low $t$ region and the remarquable agreement with the BSW prediction. 
\begin{figure}[h]
\vspace*{-3.5ex}
\begin{center}
\includegraphics[width=11cm]{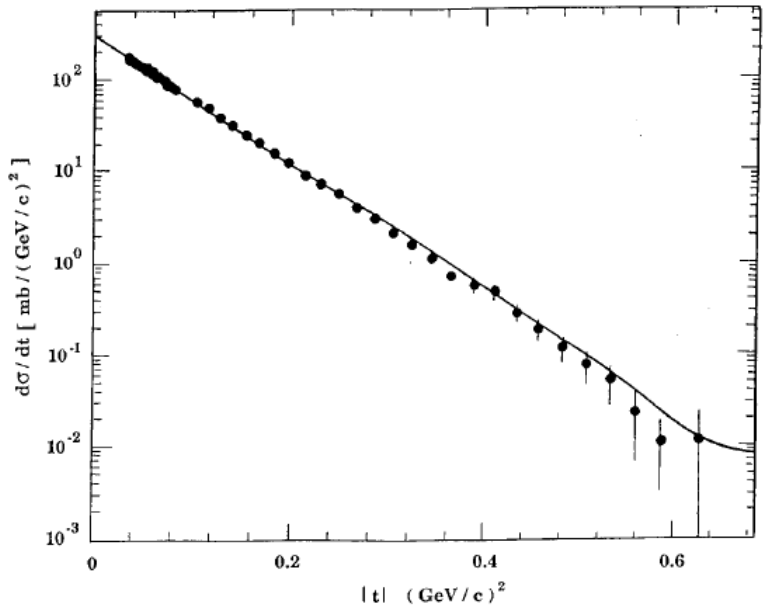}
\caption[*]{\baselineskip 1pt
 $d\sigma/dt$ for near-forward $\bar p p$ elastic scattering at  $\sqrt{s}$= 1.8TeV data from Ref. \cite{amos}, the curve is the BSW prediction \cite{bsw88} (Taken from Ref. \cite{bsw90}).}
\label{fig:dsige710}
\end{center}
\end{figure}
The BSW model predicts the correct $\rho(s)$ which appears to have a flat energy dependence in the high energy region (see Fig. \ref{fig:rho}) and for $s \to \infty$, one expects $\rho(s) \to 0$. Another specific feature of the BSW model is the fact that it incorporates the theory of expanding protons \cite{ttwu2,ttwu1}, with the physical consequence that the ratio $\sigma_{el}/\sigma_{tot}$ increase with energy. This is precisely in agreement with the data and when $s \to \infty$ one expects $\sigma_{el}/\sigma_{tot} \to 1/2$, which is the black disk limit.

Finally we show in Fig. \ref{fig:dsige710} the $t$-dependence of the elastic cross section measured by the E710 experiment, which again confirms the BSW prediction. It may be worth emphasizing that in this $t$ domain, the $t$-behavior is definitely not a straight line.
\begin{figure}[ht]
\vspace*{-8.5ex}
\begin{center}
\includegraphics[width=10.0cm]{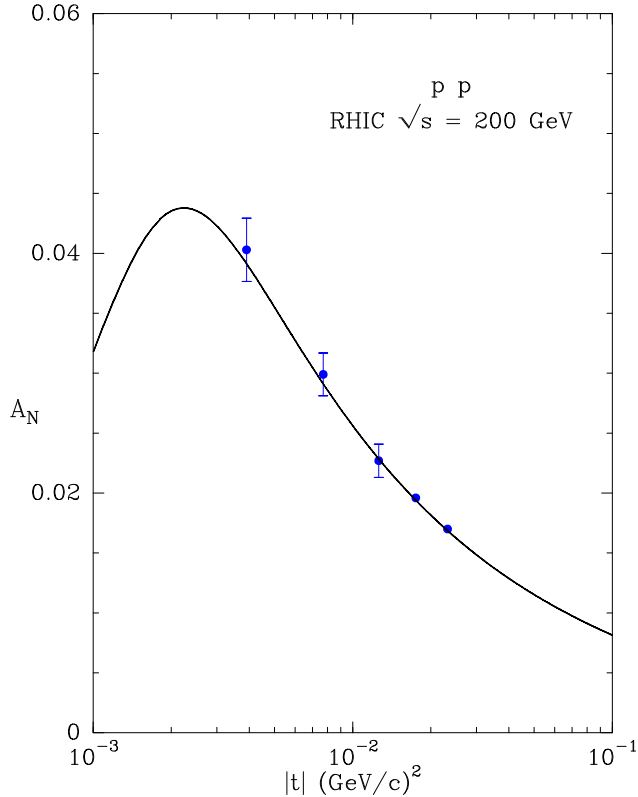}
\vspace*{-10mm}
\caption{ The analyzing 
power $A_N$ versus $t$ at RHIC energy. The data from Ref. \cite{star} are in excellent agreement with the BSW prediction.}
\label{fig:asy}
\end{center}
\end{figure}

 Before moving to the LHC energy it is worth mentioning another independent test of the BSW amplitude, by means of the analyzing power $A_N$, in $pp$
eleastic scattering near the very forward direction. In addition to the non-flip component Eq. (\ref{eq:ac}), the Coulomb amplitude has also a single-flip component $a^{C}_5$, which
involves the proton anomalous magnetic moment. In this kinematic region, the CNI region, $A_N$ results only from the interference of $a^{C}_5$, which is purely real, with the imaginary part of the hadronic non-flip amplitude, if one assumes that there is no contribution from the single-flip hadronic amplitude. This is what we have done in the calculation of the curve displayed in Fig. \ref{fig:asy} compared to some new data from STAR \cite{star}. It confirms the absence of single-flip hadronic amplitude and the right determination of $\mbox{Im}~\mathcal{M}(s, t)$ in the CNI region.

\section{ The LHC energy}

At the moment, the situation with the experimental data at the energies of
the Large Hadron Collider is not completely clear.  Here is a description
of the present data.

There are two experiments measuring the proton-proton total cross section:
TOTEM associated with the CMS detector and the ALFA associated with the
ATLAS Collaboration.  The center-of-mass energy chosen by both experiments
are 7 TeV.

The ALFA has not published yet any result from their measurements,
although such publications are expected in the very near future.  Thus the
discussion here has to be limited to those from TOTEM.  The total
proton-proton cross section given by TOTEM is
               $\sigma_{tot} (TOTEM) = (98.0 \pm 2.5)$ mb ,
which is 1.7 $\sigma$ above the BSW prediction at 7 TeV.  It should be
emphasized that the BSW parameters, shown in Table 1, was determined in
1984 and has not changed in thirty years..

\begin{figure}[hp]
\vspace*{1.0ex}
\begin{center}
\includegraphics[width=13.00cm]{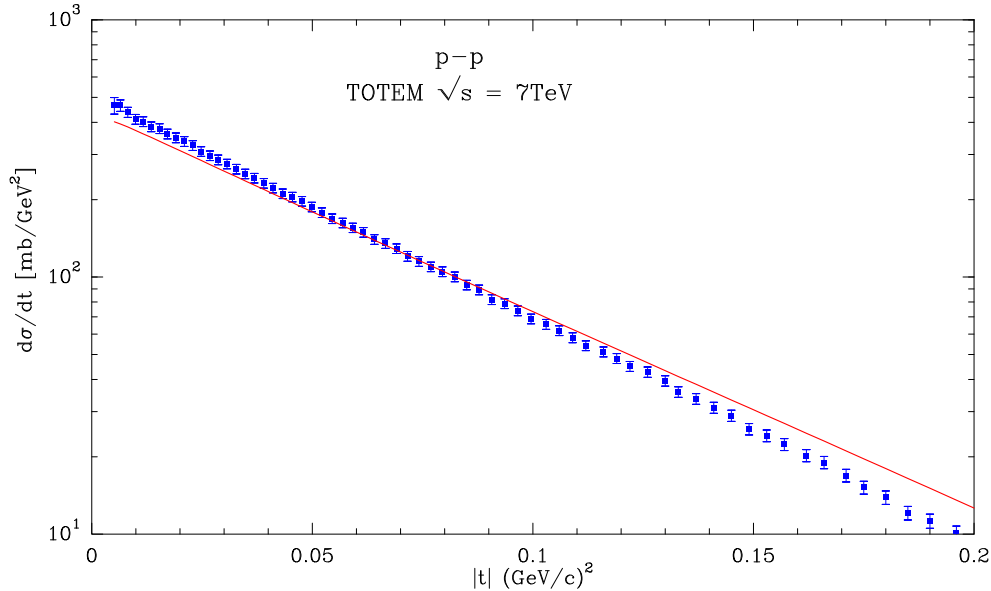}
\caption[*]{\baselineskip 1pt
$d\sigma/dt$ for near-forward $pp$ elastic scattering.
The data are from TOTEM \cite{totem} and the curve is the BSW prediction. }
\label{fig:lhc}
\end{center}
\end{figure}

As usual, the total cross section is obtained by extrapolating the
differential cross section to the forward direction.  It is therefore of
interest to compare the TOTEM measurement of the proton-proton
differential cross section directly with the BSW prediction as discussed
in Sec. 2. Such a comparison is shown in Fig. \ref{fig:lhc}.

The meaning of this comparison of Fig. \ref{fig:lhc} is not straightforward and has
not yet been completely clarified, and we look forward to future
developments, especially at even higher energies of LHC.

\section{Conclusion}

The basic idea for the development of the present phenomenology has been
described in Sec. 2.  On the one hand, it is essential to incorporate
suitably chosen results from field theory; however, some other
results from field theory have to be purposely contradicted in the
phenomenology.  Physics has played an essential role in the choices, and
is largely responsible for the success in the many predictions of the
present phenomenology, which is entirely unchanged in nearly thirty years.

When the present phenomenology was first worked out, the highest
center-of-mass energy of available experimental data was 62 GeV.  As shown
in Sec. 3, the predictions of this phenomenology are in good agreement
with later experimental data up to the center-of-mass energy of 1.96 TeV.
This is an increase of energy by a factor of more than thirty -- an
extraordinary success as discussed in the previous sections.

The conclusion is therefore reached that there is a good understanding of
proton-proton elastic scattering near the forward direction.

It will be of interest to push this phenomenology to higher energies by
comparing its predictions with the data from the Large Hadron Collider at
the center-of-mass energy of 7 TeV.  Some such comparison has already been
presented in Sec. 4, and some more data are expected from the ALFA
Collaboration in the near future.  It is even more interesting to compare
the present predictions with the data after an upgrade of the
center-of-energy of the Large Hadron Collider next year to 13 or 14 TeV,
both for the differential elastic cross section and the parameter $\rho$
discussed above, whose relevance of its measurement has been strongly emphasized \cite{bkmsw}.

It may be of some importance to add the following comment to the present
development on the increasing total cross section.  It is the production
of relatively low-energy particles in the center-of-mass system that leads
to this increasing total cross section; such production processes are
usually referred to as "pionization".  Also it was known from the
beginning \cite{ttwu2,ttwu1} that, at extremely high energies, half of this increase
in the total cross section is due to the integrated elastic cross section.
 The obvious question, already raised more than forty years ago, is: what
processes are responsible for the other half of this increase at extremely
high energies?

This question has been answered recently through the application of
geometrical optics to production processes \cite{gww}: the same processes, i.e.,
pionization, are not only responsible for the increasing total cross
section, but are also responsible for the other half of the increasing
total cross section at extremely high energies -- an answer that we find
to be philosophically satisfying.
\vskip 0.5cm
{\bf Acknowledgments}

We thank Andr\'{e} Martin and Maurice Haguenauer for several fruitful discussions.

\end{document}